\input{aipcheck}

\documentclass[
    ,final            
  ]
  {aipproc}

\layoutstyle{8x11single}

\begin{document}

\title{{\it SVOM}: a new mission for Gamma-Ray Burst Studies}


\author{D. G\"otz}{
  address={CEA Saclay - DSM/Irfu/Service d'Astrophysique - Orme des Merisiers, F-91191, Gif-sur-Yvette, France}
}

\author{J.Paul}{
  address={CEA Saclay - DSM/Irfu/Service d'Astrophysique - Orme des Merisiers, F-91191, Gif-sur-Yvette, France}
,altaddress={Astroparticules et Cosmologie (APC) - 10, rue Alice Domon et L\'eonie Duquet,  F-75205, Paris Cedex 13, France}
}

\author{S. Basa}{
  address={Laboratoire d'Astrophysique de Marseille/CNRS/Universit\'e de Provence, Technopole de l'Etoile, 38 rue Frederic Joliot-Curie, 13388 Marseille Cedex 13, France}
}

\author{J. Wei}{
  address={National Astronomical Observatories, Chinese Academy of Sciences, 20A Datun Road, Chaoyang District, Beijing 100012, China}
}

\author{S. N. Zhang}{
  address={Department of Physics and Center for Astrophysics, Tsinghua University, Beijing 100084, China}
}
\author{J.-L. Atteia}{
  address={Laboratoire d'Astrophysique de Toulouse-Tarbes, Universit\'e de
Toulouse, CNRS, 14 Av. Edouard Belin, 31400 Toulouse, France}
}

\author{D. Barret}{
  address={Centre d'Etude Spatiale des Rayonnements, Universit\'e de Toulouse, Centre Nationale de la Recherche Scientifique, 9 avenue du Colonel Roche, 31028 Toulouse Cedex 04, France}
}
\author{B. Cordier}{
  address={CEA Saclay - DSM/Irfu/Service d'Astrophysique - Orme des Merisiers, F-91191, Gif-sur-Yvette, France}
}
\author{A. Claret}{
  address={CEA Saclay - DSM/Irfu/Service d'Astrophysique - Orme des Merisiers, F-91191, Gif-sur-Yvette, France}
}
\author{J. Deng}{
  address={National Astronomical Observatories, Chinese Academy of Sciences, 20A Datun Road, Chaoyang District, Beijing 100012, China}
}
\author{X. Fan}{
  address={Xi'an Institute of Optics and Precision Mechanics, Chinese Academy of Sciences, 17 Xinxi Road, Xi'an, China}
}
\author{J.Y. Hu}{
  address={National Astronomical Observatories, Chinese Academy of Sciences, 20A Datun Road, Chaoyang District, Beijing 100012, China}
}
\author{M. Huang}{
  address={National Astronomical Observatories, Chinese Academy of Sciences, 20A Datun Road, Chaoyang District, Beijing 100012, China}
}
\author{P. Mandrou}{
  address={Centre d'Etude Spatiale des Rayonnements, Universit\'e de Toulouse, Centre Nationale de la Recherche Scientifique, 9 avenue du Colonel Roche, 31028 Toulouse Cedex 04, France}
}
\author{S. Mereghetti}{
  address={INAF--Istituto di Astrofisica Spaziale e Fisica cosmica (IASF), Via Bassini 23, I-20133, Milano, Italy}
}
\author{Y. Qiu}{
  address={National Astronomical Observatories, Chinese Academy of Sciences, 20A Datun Road, Chaoyang District, Beijing 100012, China}
}
\author{B. Wu}{
  address={Key Laboratory for particle astrophysics, Institute of High Energy Physics, Chinese Academy of Sciences, P.O.Box 918, 100049, China}
}

\begin{abstract}
We present the {\it SVOM} (Space-based multi-band astronomical Variable Object Monitor) mission, that is being developed in cooperation between the Chinese National Space Agency (CNSA), the Chinese Academy of Science (CAS)
and the French Space Agency (CNES), and is expected to be launched in 2013. Its scientific objectives include the study of the GRB phenomenon (diversity and unity), GRB physics (particle acceleration, radiation mechanisms), GRB
progenitors, cosmology (host galaxies, intervening medium, star formation history, re-ionization, cosmological parameters), and fundamental physics (origin of cosmic rays, Lorentz invariance, gravitational waves sources).
{\it SVOM} is designed to detect all known types of Gamma-Ray Bursts (GRBs), to provide fast and reliable GRB positions, to measure the broadband spectral characteristics and temporal properties of the GRB prompt emission. This will
be obtained in first place thanks to a set of four space flown instruments. A wide field ($\sim$2 sr) coded mask telescope (ECLAIRs), operating in the 4--250 keV energy range, will provide the triggers and localizations, while a gamma-ray
non-imaging spectrometer (GRM), sensitive in the 50 keV-5 MeV domain, will extend the prompt emission energy coverage. After a satellite slew, in order to place the GRB direction within field of view of the two narrow field
instruments - a soft X-ray (XIAO), and a visible telescope (VT) - the GRB position will be refined and the study of the early phases of the GRB afterglow will be possible. A set of three ground based dedicated instruments, two robotic
telescopes (GFTs) and a wide angle optical monitor (GWAC), will complement the space borne instruments.
Thanks to the low energy trigger threshold ($\sim$4 keV) of the ECLAIRs, {\it SVOM} is ideally suited for the detection of soft, hence potentially most distant, GRBs. Its observing strategy is optimized to facilitate follow-up
observations from the largest ground based facilities.
\end{abstract}

\maketitle


\section{Introduction}

Despite recent observational progress, the 40 years old Gamma-Ray Bursts (GRBs)  mystery is far from being completely
solved (see e.g. \cite{piran05} for a review). There is general consensus on the cosmological nature of these
transient sources of gamma-ray radiation, and, at least for the long burst category, the association with the
explosion of massive stars (>30 M$_{\odot}$) is a scenario that reproduces most observed features, although
the number of GRBs spectroscopically associated with Supernovae is still limited. Whatever the exact progenitors of long GRBs are, they have been detected up to redshift 6.7 \cite{fynbo08}, making them powerful tools to investigate the early Universe (star formation history, re-ionization, etc.), and to possibly derive cosmological parameters. For short bursts, on the other hand, the situation is less clear, mainly because of the lack of a statistically compelling number of good quality observations of their afterglows. Models involving the coalescence of two compact objects (black holes, neutron stars or white dwarfs) are however commonly invoked to explain these events, as their average distance is smaller with respect to long bursts. Many questions concerning
GRBs are still open, such as the physical processes at work during the prompt phase, in terms of particle acceleration
and radiation processes, the GRBs classification, a better characterization of GRB host galaxies and progenitors,
as well as some fundamental physics issues like Lorentz invariance, the origin of cosmic rays and gravitational waves.

\begin{figure}[h]
  \includegraphics[height=.2\textheight]{svom.pdf}
\hspace{2cm}
   \includegraphics[height=.2\textheight]{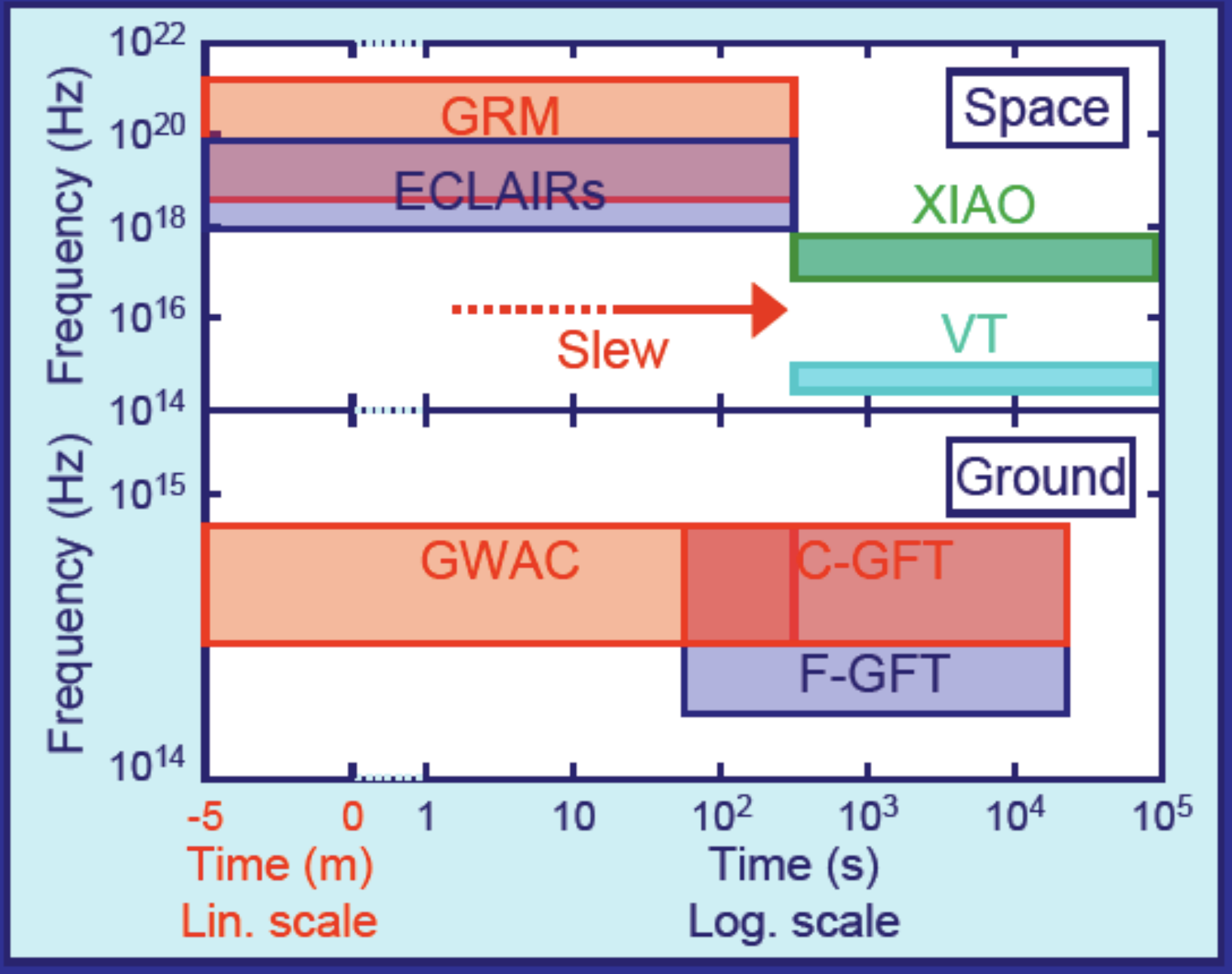}
 \caption{Left: a possible implementation of SVOM. ECLAIRs is in green, XIAO in blue, the VT in red, and the GRM units are hanging on ECLAIRs side. Right: the SVOM multi-wavelength coverage.}
\label{fig:svom}
\end{figure}

In order to contribute to address the above questions, the French Space Agency (CNES), the
Chinese Academy of Sciences (CAS) and the Chinese Space Agency (CNSA) are developing the {\it SVOM} mission 
(Space-based multi-band astronomical Variable Object Monitor). {\it SVOM} has successfully reached the end
of its phase A design study, and is planned to be launched in 2013 in a circular orbit with an inclination of 
$\sim$30$^{\circ}$ and altitude of $\sim$600 km. It will carry four instruments: ECLAIRs, a coded mask
wide field telescope that will provide real time localizations of GRB to arcminute level, two GRMs units, non-imaging gamma-ray spectrometers, and two narrow-field instruments, XIAO and VT, for arcsecond localizations, and 
for the study of the early afterglow phases in the X-ray and optical bands. Indeed, once a GRB is detected
within field of view of ECLAIRs, the satellite will autonomously perform a slew towards the GRB direction, in order
to allow the observations of the afterglow by the XIAO and VT telescopes. A possible realization of the
{\it SVOM} satellite is sketched in Fig. \ref{fig:svom}. The {\it SVOM} pointing strategy derives from a combination
of two main constraints: the avoidance of bright X-ray galactic sources and an anti-solar pointing, to have
the GRBs always detected on the night side of the Earth. Even if the latter choice induces some dead time at 
mission level, due to the Earth passages occulting ECLAIRs field of view once per orbit,
it will enhance the possibility of successful follow-up with large ground based facilities, with a goal
of 75\% of {\it SVOM} GRBs easily observable during their early afterglow phase.

Besides the space flown instruments, the {\it SVOM} mission includes a set of ground based instruments, in order
to broaden the wavelength coverage of the prompt and of the afterglow phase: GWACs are a set of wide field optical
cameras that cover a large fraction of ECLAIRs field of view. They will be based in China and will follow ECLAIRs
pointings, in order to catch the prompt optical emission associated with GRBs. Two robotic telescopes (GFTs), one based
in China, and one provided by CNES, complete the ground based instrumentation. Their goal is to measure the photometric
properties of the early afterglow from the near infra-red to the optical band, and to refine the afterglow
position provided by the on-board instruments. A summary of the multi-wavelength capabilities of SVOM
is reported in Fig .\ref{fig:svom}. In the following sections the instruments and the mission are described in some
detail.

\section{ECLAIRs}
\begin{figure}[ht]
  \includegraphics[height=.2\textheight]{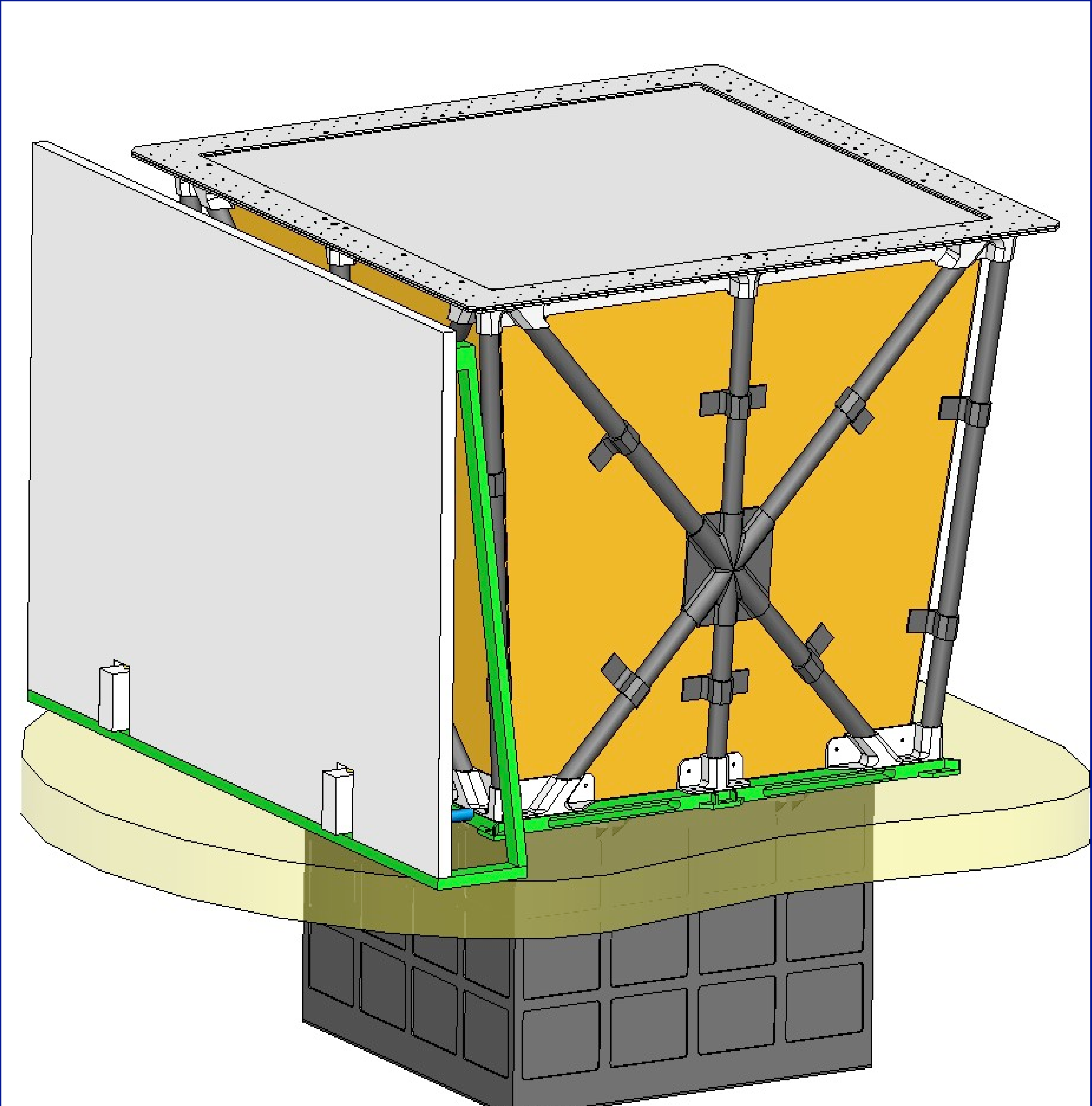}
\hspace{2cm}
  \includegraphics[height=.2\textheight]{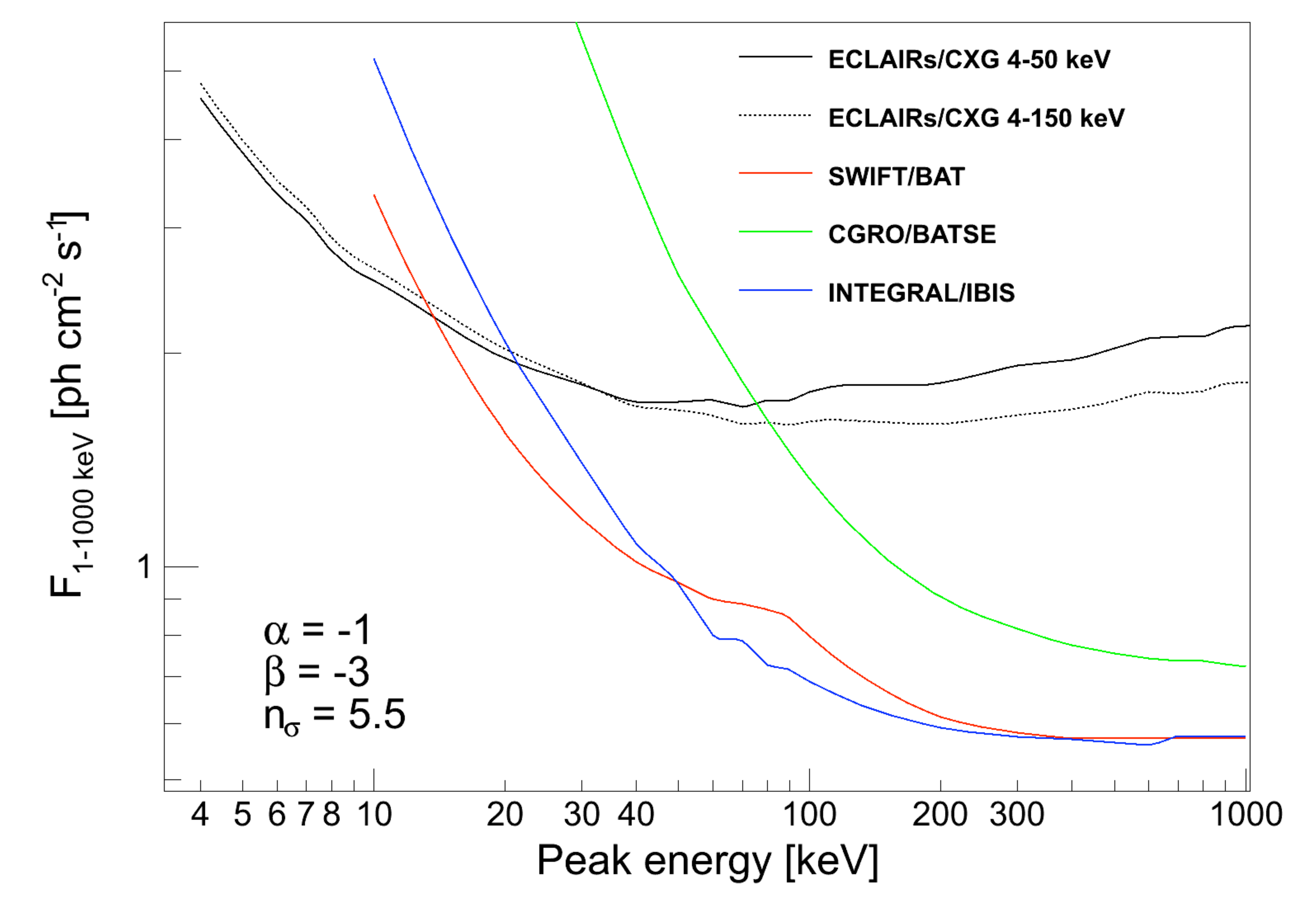}
  \caption{Left: CXG mechanical structure. The top layer represents the coded mask, the orange layers the passive shielding, and the bottom box contains the detectors electronics. Right: ECLAIRs/CXG sensitivity compared to previous and current instrumentation. The curves have been computed as a function of the GRB peak energy for a 5.5 $\sigma$ detection assuming a Band \cite{band93} spectrum with the other spectral as parameters reported in the plot.}
\label{fig:ECLAIRs}
\end{figure}

ECLAIRs is made of a coded mask telescope working in the 4--250 keV energy range (CXG), and a real-time 
data-processing electronic system (UTS, see below). The CXG has a wide field of view ($\sim$ 2 sr), and
a fair localization accuracy ($\sim$10 arcmin error radius (90\% c.l.) for the faintest sources, down to a couple of arcmin for the brightest ones).
Its detector plane is made of 80$\times$80 CdTe pixels yielding a geometrical area of 1024 cm$^{2}$. The telescope
is passively shielded, and a new generation electronics developed at CEA Saclay together with the careful detector selection, and the optimized hybridization done at CESR Toulose allow to lower the detection threshold
with respect to former CdTe detectors by about 10 keV, reaching $\sim$4 keV. The CXG, in spite
of its rather small geometrical surface, is thus more sensitive to GRBs
with soft spectra, potentially the most distant ones, than currently flying telescopes. A scheme of the CXG telescope is shown in 
left panel of Fig. \ref{fig:ECLAIRs}, while the right panel of the same figure shows how the CXG predicted sensitivity
compares to current and past missions as a function of the GRB peak energy. 

The ECLAIRs/CXG telescope is expected to localize about 70 GRBs per year. This estimate takes
into account the dead time induced by the passages over the Southern Atlantic Anomaly, that increase
significantly the instrumental particle induced background, and by the passage of the Earth in the CXG field of 
view.

\subsection{UTS}
The UTS (Unit\'e de Traitement Scientifique) is in charge of analyzing 
ECLAIRs data stream in real-time and of detecting and localizing the GRBs occurring within its field of view. It will implement two separate
triggering algorithms, one based on the detection of excesses in the detectors count rate, and using
imaging as trigger confirmation and localization, and a second one based on imaging, that is better
suited for long, slowly rising GRBs. Once a GRB is detected, its position will be sent to the ground via
a VHF emitter antenna. At the same time the GRB position will be transmitted to the platform that may
autonomously slew to the GRB direction in three to five minutes  in order to bring it in the field of view of the narrow field instruments, XIAO and VT. 
The UTS will also collect relevant information in the CXG and GRM data stream, such as light curves
and sub-images, and send them to the Earth via the VHF channel, to promptly provide additional information 
on the trigger quality and on the GRB spectral and temporal characteristics. For more details on ECLAIRs, see \cite{schanne08} and \cite{remoue08}. 

\section{GRM}

\begin{figure}[ht]
 \includegraphics[height=.2\textheight]{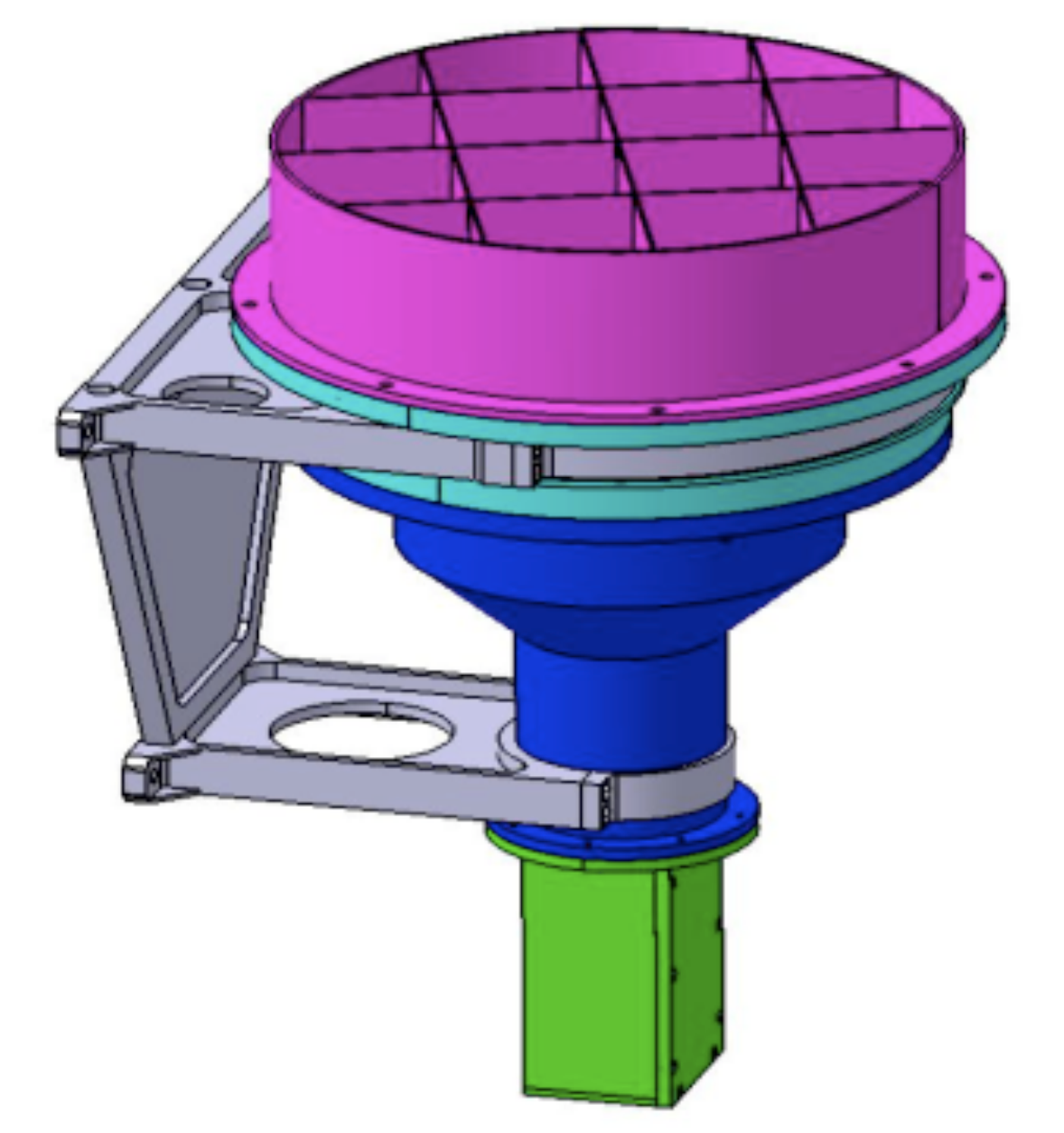}
\hspace{2cm}
 \includegraphics[height=.3\textheight]{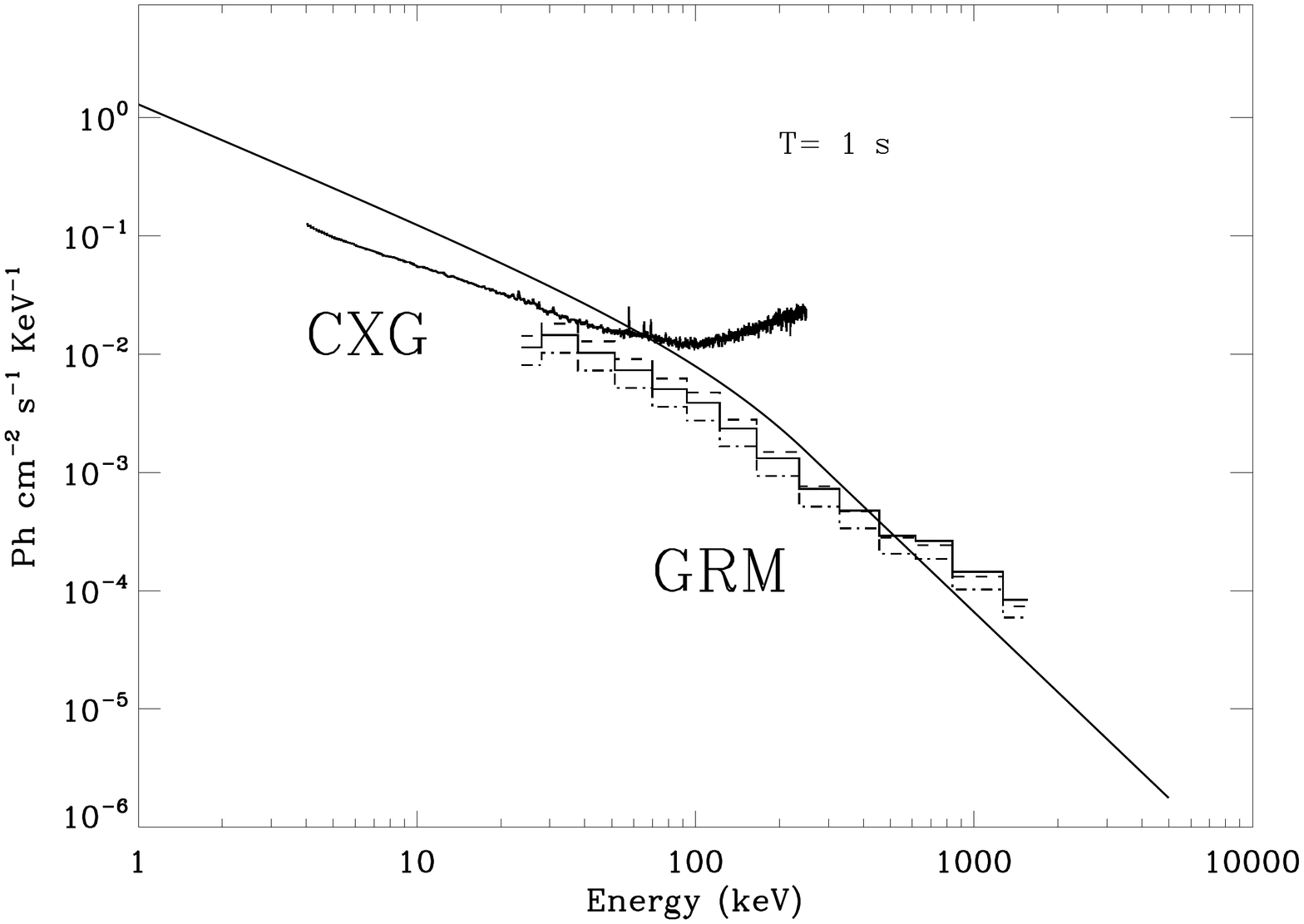}
  \caption{Left: one GRM unit. Right: combined GRM (two units) ECLAIRs on axis sensitivity for a 1 s integration time and a 5 $\sigma$ detection. A Band spectrum with $\alpha$=1, $\beta$=2.5, $E_{0}$=100 keV, and F$_{50-300 keV}$=1 photon cm$^{-2}$ s$^{-1}$ is overplotted for comparison.}
\label{fig:GRM}
\end{figure}

The Gamma-Ray Monitor (GRM) on board {\it SVOM} is composed of two identical units each made of a phoswich (NaI/CsI) detector of 280 cm$^{2}$, read by a photomultiplier. In front of each detector there is a collimator in order to reduce the background and to match the CXG and GRM fields of view. A schematic view of one GRM unit and the combined
CXG/GRM sensitivity are shown in Fig. \ref{fig:GRM}. The GRM does not have imaging capabilities, however as can be seen from Fig. \ref{fig:GRM}, the GRM extends the spectral coverage of the {\it SVOM} satellite to the MeV range. This
is an important point, since the current detectors like BAT on board {\it Swift} \cite{swift} or IBIS/ISGRI on board {\it INTEGRAL} \cite{isgri} have a comparable or better localization accuracy with respect to the CXG, but they lack a broad band coverage, hampering a correct spectral characterization of the prompt high-energy emission of GRBs, which is a key input
of any sensible modeling of GRBs' radiative processes. The recent successful launch of {\it Fermi}, and the availability
of the GBM detectors (non-imaging spectrometers with a 2$\pi$ field of view) used in synergy with BAT and ISGRI will
partially fill this need before the launch of {\it SVOM}, but the different orbits, pointing constrains, and sensitivities
of the three instruments imply a low rate of simultaneous detections. On the other hand, for every {\it SVOM} GRB a good
localization and good spectral information will be available at the same time. 

\section{XIAO}

\begin{figure}[ht]
 \includegraphics[height=.25\textheight]{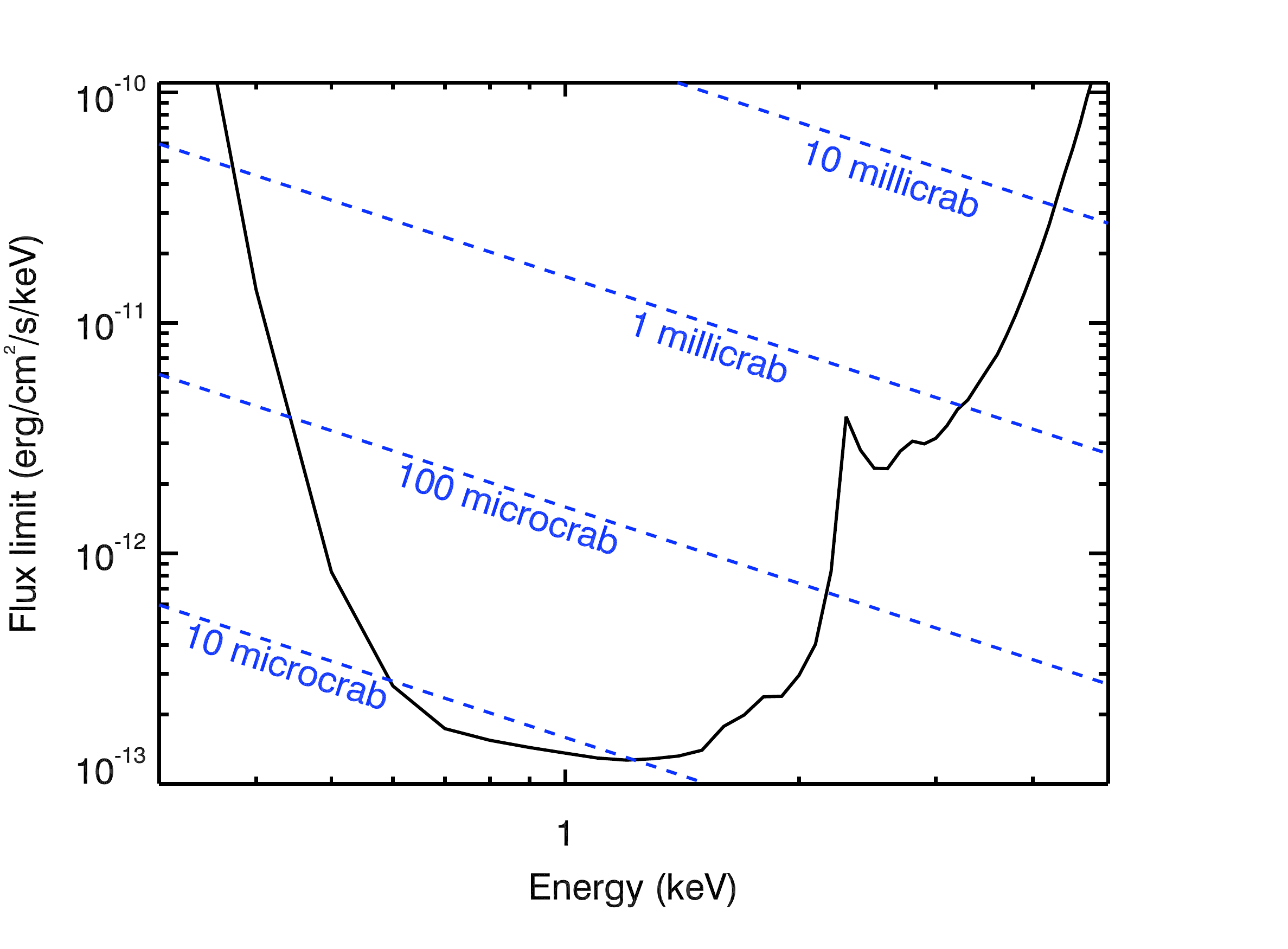}
 \includegraphics[height=.25\textheight]{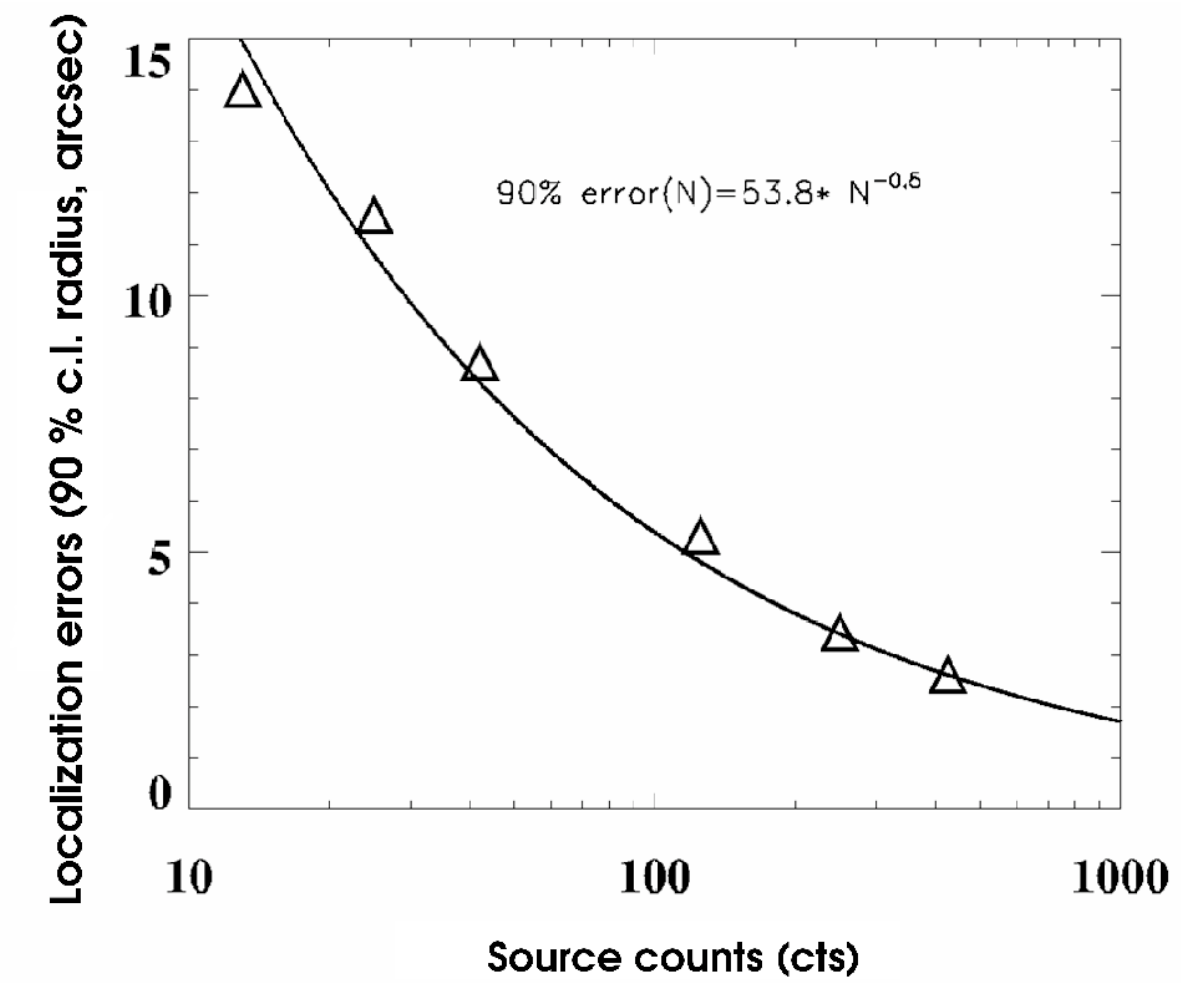}
  \caption{Left: Expected sensitivity of the XIAO telescope (5$\sigma$ detection in 10 ks), computed assuming a 30$^{\prime\prime}$ Half Energy Diameter, a source extraction circle of 15$^{\prime\prime}$ radius, and an energy bin $\Delta$E=E/2. Right: XIAO location accuracy as a function of the number of detected photons. For each afterglow at least 100 counts are expected to be collected by XIAO.}
\label{fig:XIAO}
\end{figure}

The X-ray Imager for Afterglow Observations (XIAO) will be provided by an Italian consortium, lead
by the INAF-IASF institute in Milan. It is a focusing X-ray telescope, based on the grazing incidence (Wolter-1) technique.  It has a short focal length of $\sim$0.8 m, and a field of view  (25 arcmin diameter) adequate to cover the whole error region provided by the CXG telescope, so that after the satellite slew the GRB position should always be inside the XIAO field of view. XIAO has an effective area of about 120 cm$^{2}$ and the mirrors are coupled to a very compact, low noise, fast read out CCD camera, sensitive in the 0.5--2 keV energy range. The sensitivity of the XIAO telescope is reported
in Fig. \ref{fig:XIAO}, and simulations based on a sample of light curves collected with XRT
on board {\it Swift} show that virtually all GRB X-ray afterglows are detectable by XIAO during the first hours. This means  in practice that each GRB, for which a satellite slew is performed (not all GRB can be pointed due to different
constraints at platform level), can be localized with a $\sim$ 5 arcsec accuracy, see Fig. \ref{fig:XIAO}. Indeed the source
localization accuracy, is linked to the number of detected photons as $k / \sqrt{N}$ where k is a constant depending on the instrument point spread function, and N is the number of detected photons, and the early afterglows will provide
enough photons to reach the degree of positional accuracy mentioned above.
For more details on the XIAO instrument, see \cite{mereghetti08}

\section{VT}
\begin{figure}[ht]
 \includegraphics[height=.2\textheight]{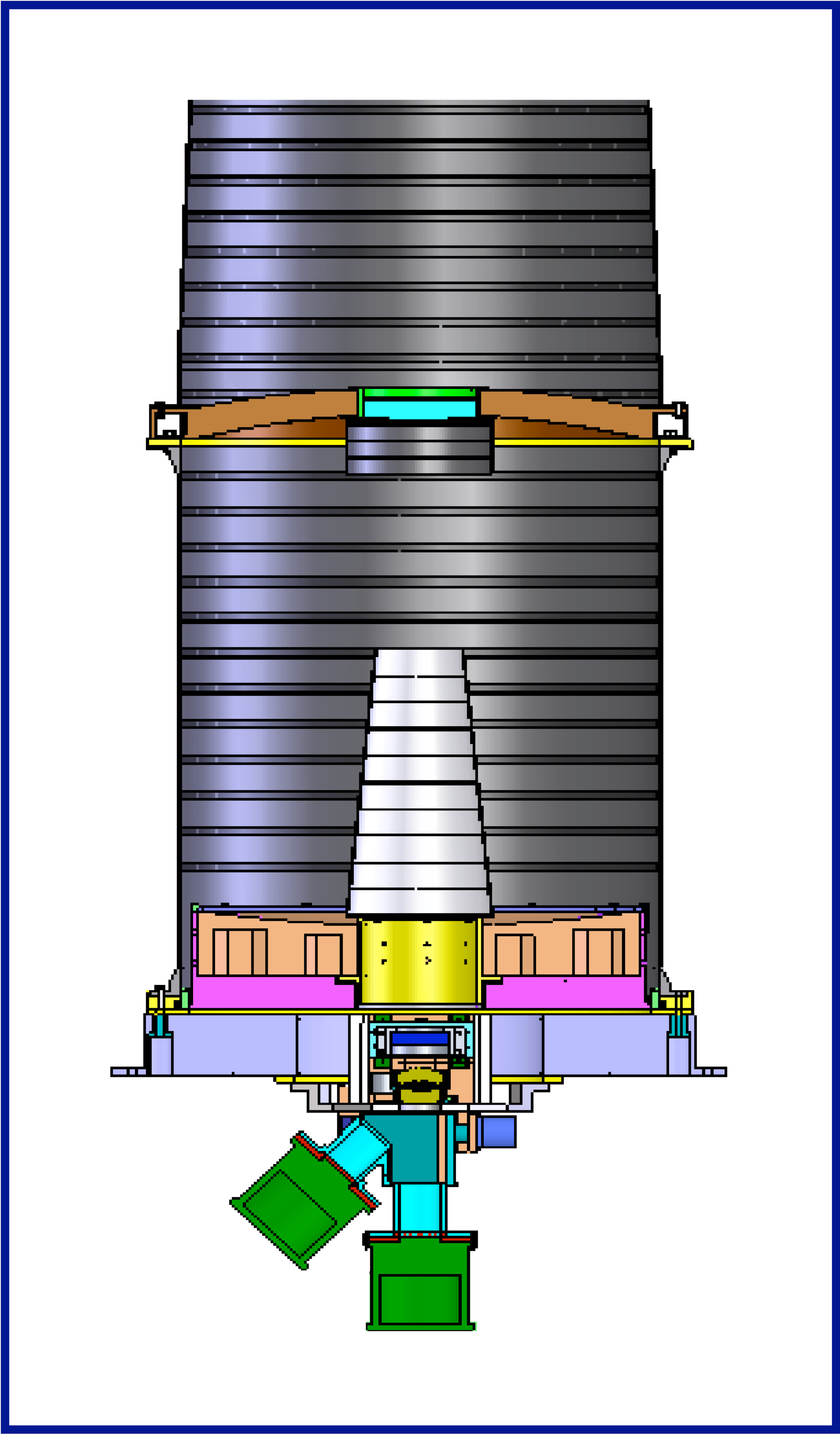}
 \hspace{2cm}
\includegraphics[height=.2\textheight]{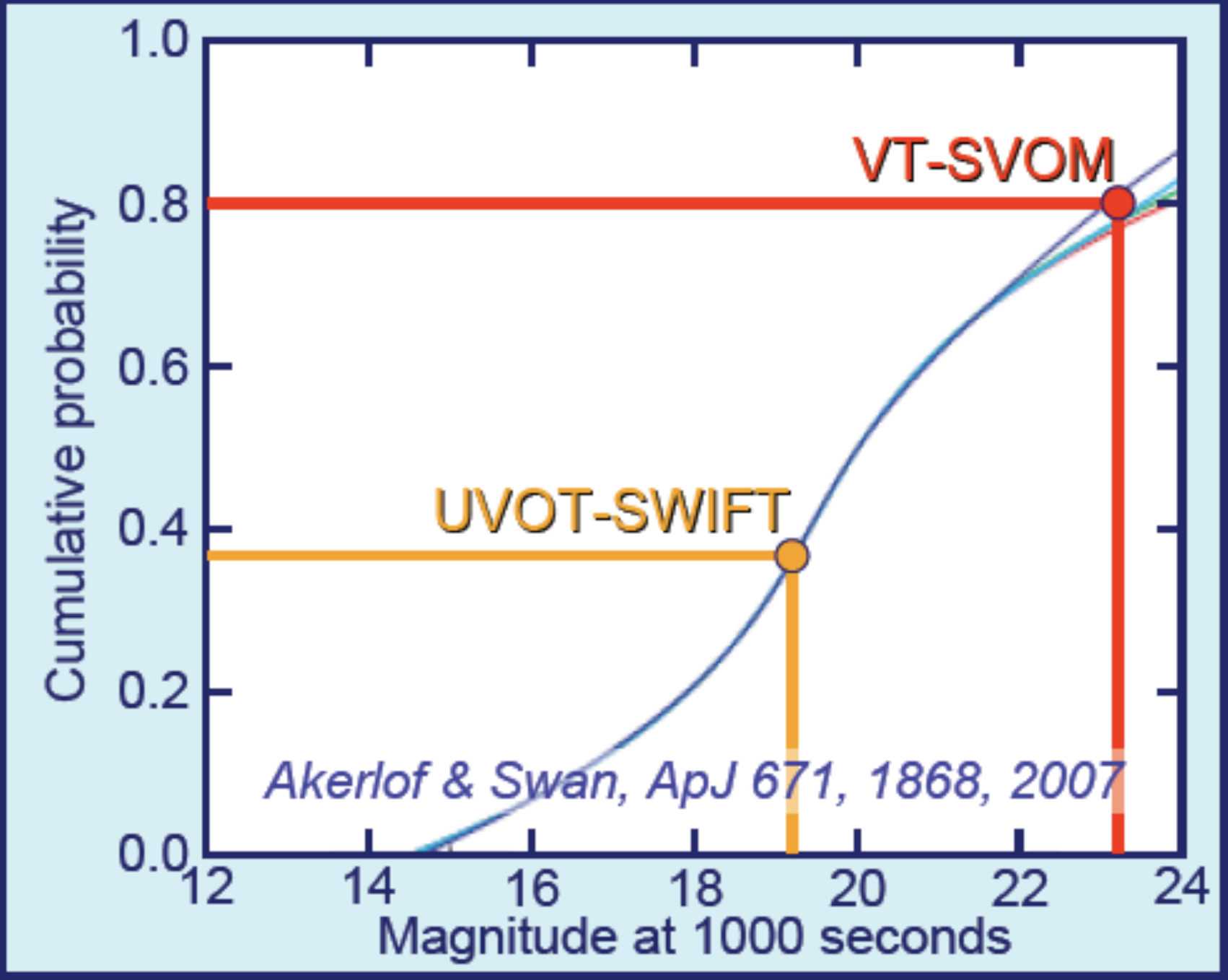}
  \caption{Left: VT. Right: Predicted optical afterglow light curves compared to UVOT and VT sensitivities in 1000 s.}
\label{fig:vt}
\end{figure}

The space-borne Visible Telescope will be able to improve the GRB localizations obtained by the CXG and XIAO to sub-arcsecond precision through the observation of the optical afterglow. In addition it will provide
a deep and uniform light-curve sample of the detected optical afterglows, and allow
to do primary selection of optically dark GRBs and high-redshift GRB candidates (z>4).
The field of view of the telescope will be 21$\times$21 arcmin, sufficient to cover the error box of the CXG. The detecting area of the CCD has 2048 x 2048 pixels to ensure the sub-arcsecond localization of detected sources.
The aperture of the telescope should guarantee a limiting magnitude of $M_{V}$ = 23 (5$\sigma$) for a 300 s  exposure time. Such a sensitivity is a significant improvement over the UVOT on board the {\it Swift} satellite (see Fig. \ref{fig:vt}) and over existing ground-based robotic GRB follow-up telescopes. The VT is expected  to detect nearly 70\% of SVOM GRBs for which a slew is performed.
The telescope will have at least two bands in order to select high-redshift GRB candidates. They are separated at 650 nm, which corresponds to a redshift of z$\sim$4-4.5 using Ly$\alpha$ absorption as the redshift indicator.

\section{Ground segment}
The ground segment of the mission will be composed of X- and S-band antennas (for data and housekeeping telemetry
download), a mission operation center, based in China, two science centers (based in China (CSC) and France (FSC) and in charge of operations and monitoring of the scientific payload), 
and a VHF alert network.The latter will be composed of a series of receivers distributed over the 
globe in order to guarantee continuous coverage for the alerts dispatched by the platform. The alerts
will contain the information about the GRB positions, that will be sent to the ground as soon as more
accurate information is derived on board, followed by complementary quality indicators (light curves, images,
etc.) produced on board. The VHF network is directly connected to the FSC, which is in charge of 
formatting and dispatching the alerts to the scientific community through the Internet (GCNs, VO Events, SVOM web page, etc.). The first alerts corresponding to the initial localization by the CXG are expected to reach the recipients
one minute after the position has been derived on board. Then the following alerts, containing the X-ray
position of the afterglow, and eventually the optical position computed from VT data, will reach the community
within 10 minutes from the first notice. In case of a refined (sub-arcsec) position is available from the prompt
data analysis of the GFTs or GWACS (see below), this information will immediately reach the the FSC, and dispatched to the 
scientific community through the channels mentioned above. For more details on the alert distribution strategy, see \cite{claret08}. In addition the FSC will be on charge of publishing the CXG pointing direction in order to facilitate
ground based robotic telescopes to quickly react, minimizing the slew time.


\subsection{Ground Based Telescopes}
\subsubsection{GWACS}
The Ground-based Wide-Angle Camera array is designed to observe the visible emission of more than 20\% of 
{\it SVOM} GRBs from 5 minutes before to 15 minutes after the GRB onset.  The array is expected to have an assembled field of view of about 8000 deg$^{2}$ and a 5$\sigma$ limiting magnitude $M_{V}$ = 15 for a 15 s exposure time for full moon nights. To comply with both the science requirements and technical feasibility, each camera unit will have an aperture size of 15 cm, a 2048 x 2048 CCD and a field of view of 60 deg$^{2}$. In total about 128 camera units are required to cover the 8000 deg$^{2}$.
\subsubsection{GFTs} 
Two SVOM robotic telescopes will automatically position their 20-30 arcmin field-of-view to the position of GRB alerts and, in case of a detection, they will determine the position of the source with 0.5 arcsec accuracy. Both telescopes will be provided with multi-band optical cameras and the French GFT will also have a near infra-red CCD.
Since one telescope can only observe those candidates occurring above the horizon and during night at the telescope site,  these sites will be located in tropical zones and at longitudes separated by 120$^{\circ}$ at least, in order to fulfill the requirement of a 40\% efficiency. These telescopes could also be used to follow alerts which are not considered to be reliable enough to be distributed to the whole community. This procedure allows increasing the chance of detecting low S/N events, while not wasting the observing time of instruments which are outside the {\it SVOM} collaboration. 

The scientific objectives of the GFTs include the quick identification and characterization of interesting GRBs 
(e.g. highly redshifted GRBs, whose visible emission is absorbed by the Lyman alpha cutoff and the Lyman alpha forest, dark bursts, nearby GRBs), and multi-wavelength follow-up of 40\% of {\it SVOM} GRBs (at optical and X-ray wavelengths) from 30 to 10$^{4}$ seconds after the trigger. This will be done with the GFT and the VT at optical wavelengths and CXG and the XIAO at X-ray wavelengths. This will allow measuring the spectral energy distribution of the burst during the critical transition between the GRB and the afterglow.




\begin{theacknowledgments}
D.G. acknowledges the French Space Agency (CNES) for financial support. J.D. \& Y.Q. are supported by NSFC (No. 10673014). S.M. acknowledges the support of the Italian Space Agency through contract I/022/08/0.
\end{theacknowledgments}



\bibliographystyle{aipproc}   

\end{document}